\documentclass[aps,prl,twocolumn,nofootinbib,preprintnumbers]{revtex4}  % for review and submission
\usepackage{graphicx}  % needed for figures
\usepackage{dcolumn}   % needed for some tables
\usepackage{bm}        % for math
\usepackage{amssymb}   % for math
\usepackage{xcolor}
\def\appendix{{\newpage\section*{Appendix}}\let\appendix\section%
        {\setcounter{section}{0}
        \gdef\thesection{\Alph{section}}}\section}

\newcommand\ba{\begin{eqnarray}}
\newcommand\ea{\end{eqnarray}}

\definecolor{DarkGreen}{rgb}{0,.64,0}
\definecolor{gunmetal}{rgb}{0.171875, 0.207031, 0.222656}
\definecolor{chartreuse}{rgb}{.49,.98,0}
\definecolor{amethyst}{rgb}{0.59375,0.398438,0.792969}
\definecolor{brownrust}{rgb}{0.6875, 0.316406, 0.242188}
\definecolor{Violet}{rgb}{0.5,0,1}
\definecolor{BurntOrange}{rgb}{0.792969,0.332031,0}
\definecolor{FreshEggplant}{rgb}{0.59375, 0., 0.414063}
\definecolor{salmon}{rgb}{0.996094,0.507813,0.410156}
 \definecolor{FrenchRose}{rgb}{0.96875, 0.292969, 0.5625}
\definecolor{Cabaret}{rgb}{0.808594, 0.242188, 0.46875}
\definecolor{Shamrock}{rgb}{0.242188, 0.808594, 0.582031}
\definecolor{RobinsEggBlue}{rgb}{0., 0.792969, 0.792969}
\definecolor{GuardsmanRed}{rgb}{0.792969, 0., 0.}
\definecolor{Sapphire}{rgb}{0.183594, 0.328125, 0.621094}
\definecolor{Sorbus}{rgb}{0.996094, 0.429688, 0.0273438}
\definecolor{Red}{rgb}{1,0,0}
\definecolor{Blue}{rgb}{0,0,1}
\definecolor{Green}{rgb}{0,1,0}
\definecolor{thistle3}{rgb}{0.800781, 0.707031, 0.800781}
\definecolor{thistle4}{rgb}{0.542969, 0.480469, 0.542969}

\newcommand{\BREVITIZEnoA}[1]{}
\newcommand{\BREVITIZEnoB}[1]{}
\newcommand{\BREVITIZEnoC}[1]{}

\newcommand{\BREVITIZEnoE}[1]{}

\newcommand{\SilentGreen}[1]{}

\newcommand{\shgMUTE}[1]{}

\newcommand{\nn}{\nonumber}

\def\Dslash{\,\,{\raise.15ex\hbox{/}\mkern-12mu D}}
\def\Dbarslash{\,\,{\raise.15ex\hbox{/}\mkern-12mu {\bar D}}}
\def\delslash{\,\,{\raise.15ex\hbox{/}\mkern-9mu \partial}}
\def\delbarslash{\,\,{\raise.15ex\hbox{/}\mkern-9mu {\bar\partial}}}
\def\pslash{\,\,{\raise.15ex\hbox{/}\mkern-9mu p}}
\def\calDslash{\,\,{\raise.15ex\hbox{/}\mkern-12mu {\cal D}}}

\newcommand{\hh}{{1\over 2}}

\renewcommand{\ll}{_}
\newcommand{\uu}{^}
\newcommand{\pp}{\partial}

\newcommand{\expr}[1]{{\rm exp}\left ( #1 \right ) }
\renewcommand{\d}{\delta}
\newcommand{\m}{\mu}

\renewcommand{\m}{\mu}
\newcommand{\n}{\nu}
\newcommand{\s}{\sigma}

\renewcommand{\a}{\alpha}

\renewcommand{\o}{\omega}
\newcommand{\e}{\epsilon}

\newcommand{\sqd}{^2}

\renewcommand{\hh}{{1\over 2}}

\newcommand{\eee}[1]{\ba{#1}\ea}

\renewcommand{\b}{\beta}

\newcommand{\apr}{{\alpha^\prime} {}}

\newcommand{\IZ}{\relax\ifmmode\mathchoice
{\hbox{\cmss Z\kern-.4em Z}}{\hbox{\cmss Z\kern-.4em Z}}
{\lower.9pt\hbox{\cmsss Z\kern-.4em Z}} {\lower1.2pt\hbox{\cmsss
Z\kern-.4em Z}}\else{\cmss Z\kern-.4em Z}\fi} \font\cmss=cmss10
\font\cmsss=cmss10 at 7pt
\newcommand{\inbar}{\,\vrule height1.5ex width.4pt depth0pt}
\newcommand{\IC}{{\relax\hbox{$\inbar\kern-.3em{\rm C}$}}}
\newcommand{\IQ}{{\relax\hbox{$\inbar\kern-.3em{\rm Q}$}}}
\newcommand{\IP}{\relax{\rm I\kern-.18em P}}

\newcommand{\ed}{\dot{e}}

\newcommand{\cc}{{\cal C}}

\renewcommand{\cc}{{c_1}}

\renewcommand{\o}{\omega}

\renewcommand{\cc}{c}

\newcommand{\IR}{\relax{\rm I\kern-.18em R}}
\def\blfootnote{\xdef\@thefnmark{}\@footnotetext}

\renewcommand{\cc}[1]{\cite{#1}}

\newcommand{\ee}[1]{\ba {#1} \ea}

\newcommand{\upp}[1]{^{({#1})}}

\newcommand{\rr}[1]{(\ref{{#1}})}

\newcommand{\bbb}{\ba}
\renewcommand{\eee}{\ea}
\newcommand{\een}[1]{\label{#1}\ea}
\def\xxn{\nn\\{} }

\newcommand{\kket}[1]{\left | {#1} \right \rangle }
\newcommand{\bbra}[1]{\left \langle {#1} \right | }

\def\bi{\begin{itemize}}
\def\ei{\end{itemize}}

\def\ed{\end{document}}

\def\cc{{\cal C}}

\renewcommand{\rr}[1]{(\ref{#1})}

\def\cc{}

\def\cc{\,}

\def\Zb{{\bar{Z}}}

\newcommand{\lp}[1]{_{({#1})}}

\usepackage{fancyhdr}
\usepackage{dsfont}

\newcommand{\outt}[1]{}

\renewcommand{\bm}{\begin{matrix}}
\renewcommand{\em}{\end{matrix}}

\newcommand{\aaa}[1]{}

\def\be{\begin{eqnarray}}
\def\ee{\end{eqnarray}}
\def\nn{\nonumber}

\newcommand{\WhiteOutWithNotification}[1]{ {  \color{Red}  (  A SECTION HAS BEEN WHITED OUT HERE. \cc   )  } }
\newcommand{\WhiteOut}[1]{}

\def\Zb{{\bar{Z}}}

\newcommand{\ddcoket}[1]{ \Big| \cc [{#1}]  \cc   \Big\rangle\kern-5pt \Big\rangle }
\newcommand{\ddcobra}[1]{ \Big\langle\kern-5pt \Big\langle \cc [{#1}]  \cc  \Big|}
\newcommand{\ddcoin}[2]{\Big\langle \kern-5pt \Big \langle \cc [{#1}] \cc \Big |  \cc [{#2}]  \cc  \Big \rangle \kern-5pt \Big \rangle }
\newcommand{\ddcoketinline}[1]{ \big| \cc [{#1}]  \cc   \big\rangle\kern-3pt \big\rangle }
\newcommand{\ddcobrainline}[1]{ \big\langle\kern-3pt \big\langle \cc [{#1}]  \cc  \big|}
\newcommand{\ddcoininline}[2]{\big\langle \kern-3pt \big \langle \cc [{#1}] \cc \big |  \cc [{#2}]  \cc  \big \rangle \kern-3pt \big \rangle }

\def\redlowdash{{\color{Red}{\rule[-0.5ex]{2pt}{0.4pt}}}}
\def\redmiddash{{\color{Red}{\rule[+0.5ex]{2pt}{0.4pt}}}}

\def\cute{{\lower3.5pt\hbox{\sixly
  \kern-.21pt \char58 \kern-.21pt }}}

\def\midcute{{\lower-1.0pt\hbox{\sixly
  \kern-.21pt \char58 \kern-.21pt }}}

  \def\lowcute{{\lower3.5pt\hbox{\sixly
  \kern-.21pt \char58 \kern-.21pt }}}
  
  \def\redmidcute{{\color{Red} \midcute}}
  
  \def\redlowcute{{\color{Red} \lowcute}}
   
    \def\bluelowcute{{\color{Blue} \lowcute}}

   \def\swave{\bgroup \markoverwith \midcute \ULon} 
  
  \def\redswave{\bgroup \markoverwith \redmidcute \ULon} 
  
  \def\reduline{\bgroup \markoverwith \redlowdash \ULon}
  
   \def\blueuline{\bgroup \markoverwith \bluelowdash \ULon}
  
   \def\reduwave{\bgroup \markoverwith \redlowcute \ULon}
   
   \def\blueuwave{\bgroup \markoverwith \bluelowcute \ULon}
  
  \def\redsout{\bgroup \markoverwith \redmiddash \ULon}
  
   \def\bluesout{\bgroup \markoverwith \bluemiddash \ULon}

\newcommand{\isgCOBI}[1]{{}}
\newcommand{\greenCOBICUT}[1]{{}}

\newcommand{\ShortFastCUT}[1]{{}}
\begin{document}
\preprint{IPMU13-0230}

% The following information is for internal review, please remove them for submission
%\widetext
%\leftline{Version xx as of \today}
%\leftline{Primary authors: Joe E. Physics}
%\leftline{To be submitted to (PRL, PRD-RC, PRD, PLB; choose one.)}
%\leftline{Comment to {\tt d0-run2eb-nnn@fnal.gov} by xxx, yyy}
%\centerline{\em D\O\ INTERNAL DOCUMENT -- NOT FOR PUBLIC DISTRIBUTION}

% the following line is for submission, including submission to the arXiv!!
%\hspace{5.2in} \mbox{Fermilab-Pub-04/xxx-E}

\title{String Theory of the Regge Intercept}
\author{S. Hellerman}
\email{simeon.hellerman.1@gmail.com}
\affiliation{Institute for the Physics and Mathematics of the Universe\\
The University of Tokyo \\
 Kashiwa, Chiba  277-8582, Japan}

\author{I. Swanson}
\email{ianswanson.physics@gmail.com}
\noaffiliation

\date{\today}

\begin{abstract}
Using the Polchinski-Strominger 
effective string theory in covariant gauge, we 
compute the mass of a rotating string in $D$ dimensions with large
angular momenta $J$, in one or two planes, in fixed ratio,
up to and including first subleading order in the large $J$ expansion.  
This constitutes a first-principles calculation of the value for the 
order-$J\uu 0$ contribution to the mass-squared of a meson on the 
leading Regge trajectory in planar QCD with bosonic quarks.  For open strings with Neumann boundary conditions, and
for closed strings in $D\geq 5$, the order-$J\uu 0$ term in the 
mass-squared is exactly calculated by the semiclassical
approximation. This term in the expansion is universal and independent of the details
of the theory, assuming only $D$-dimensional Poincar\'e invariance and the absence of other 
infinite-range excitations on the string worldvolume, beyond the Nambu-Goldstone bosons.
\end{abstract}

\maketitle

\section{Introduction}
The dynamics of relativistic strings
were first studied as a model to explain the observed spectra of hadronic
resonances, which are organized in families according to the mass relation 
\bbb
M\sqd \simeq \frac{ J}{\apr} + M\ll 0\sqd = {{J - \a\ll 0}\over {\apr}}\ ,
\een{RegSpec}
where $J$ is the angular momentum, $\apr$ defines the Regge slope, and the Regge intercept 
$\a\ll 0 = - \apr\cc M\ll 0 \sqd $ depends on the family.
In the string model, the Regge slope can be expressed in terms of the string tension
${\bf T}\ll{\rm string}$, with $\apr = {1}/({2\pi {\bf T}\ll{\rm
string}})$.  Empirically, the string tension takes the value ${\bf T}\ll{\rm string} \simeq 0.17~ {\rm Gev}\sqd$, 
as determined, for instance, from the spectra of quarkonia 
(see, e.g., \cite{bgtRegge, btRegge, someExperimentalResults, PDG}).

While the string model successfully describes the linear dependence of $M\sqd$ on $J$ in terms of 
an underlying relativistic (classical) dynamics, it has long been unclear how to
promote the string theory of quantum chromodynamics from
a coarse phenomenology to a precision 
science.   In particular, one would like to be able to view
eqn.~\rr{RegSpec} as capturing the leading terms in an asymptotic expansion that
holds at large angular momentum $J$.  The detailed features of Regge spectra in planar confining
theories, beyond the linear dependence on angular
momentum, have received relatively little attention from string
theorists, in part becuase the simplest quantizations of the string involve new light degrees of freedom
on the worldvolume, rather than a minimal theory of the embedding
coordinates of the string alone.  
As a result, quantization of the string worldsheet has thus far been unable to promote the classical picture of
the string to a controlled
approximation at large $J$ with calculable and universal corrections.

In this letter we quantize the effective theory of a rotating relativistic string
in $D$ dimensions in conformal gauge, and we calculate the energy of a
string with large angular momentum $J$, in one or two planes, in
  fixed ratio, up to and including the first
subleading order in the $1/J$
expansion.  That is, we calculate the order-$|J|\uu 0$ term in the mass-squared of a rotating
string on the lowest-mass Regge trajectory at large $J$.

We perform our calculation in the spirit and in the formalism of
\cite{Polchinski:1991ax}, wherein Polchinski and Strominger proposed a unitary quantization of 
the relativistic string that preserves Lorentz invariance at the quantum level without introducing degrees of freedom
beyond the motions of the string itself in $D$-dimensional space-time. 
This approach, known as ``effective string theory," can be implemented in any dimension $D$, and
can describe the spacetime kinematics of strings without any additional degrees of freedom.
This is similar to the Polyakov approach \cite{polyakov}, which couples a conformal
field theory describing $D$ free embedding coordinates $X\uu\m$ to an intrinsic worldsheet metric
$g\ll{ab}$, and treats worldsheet reparameterizations and Weyl transformations of $g\ll{ab}$ as gauge symmetries.  
The Polchinski-Strominger (PS) approach differs by introducing $X\uu\m$ 
variables that are not free, such that the central charge is compensated by
interaction terms.  In particular, in addition to the free Lagrangian
\bbb
{\cal L}\ll{\rm free} = {1\over{\pi\apr}} \cc \pp\ll + X\cdot\pp\ll - X\ ,
\een{Lfree}
the theory has an interaction term
\bbb
{\cal L}_{\rm PS} =
 {\frac{\beta}{2\pi}} {\frac{ ( \pp\ll + \sqd X \cdot \pp\ll - X ) ( \pp\ll +  X \cdot \pp\ll -\sqd  X )  }
			{(\pp\ll + X \cdot \pp\ll - X)\sqd  }}\ ,
\een{psterm}
which controls the conformal anomaly by contributing $\Delta c = 12 \b$   
to the central charge of the conformal dynamics of the $X\uu\m$
coordinates.  (In the above, we use 
worldsheet coordinates $\s\uu \pm \equiv \s\uu 0 \pm \s\uu 1$.)
Though the Polyakov formalism is not the starting point, the resulting action and constraints
are exactly the same as if we had coupled the theory ${\cal L} = {\cal
  L}\ll{\rm free} + {\cal L}_{\rm PS}$ to an
intrinsic metric, and then gauge-fixed to a flat metric $g\ll{ab} = - \hh(\d\ll{a+} \d\ll{b-} + \d\ll{a-} \d\ll{b+})$,
with $\b$ chosen so that the theory of the $X\uu\m$ is an interacting CFT with $c=26$.
This fixes the value
\bbb
\b = {{26-D}\over{12}}\ .
\een{betaval}
The action may be supplemented
with terms of order $|X|\uu{-2}$ or smaller, as consistent with (or required by) the condition that conformal invariance be
maintained order by order in $|X|$.   These terms contribute to
amplitudes of order-$J\uu{-2}$ and smaller, relative to leading-order
quantities, and therefore do not contribute to the asymptotic Regge intercept.

The interacting stress tensor of the $X$ fields satisfies the OPE of a conformal stress tensor with
$c=26$ \cite{Polchinski:1991ax}; its modes are Virasoro generators, %satisfying the Virasoro algebra and defining
defining physical states $\kket\Psi$ under the conditions 
\bbb
(L\ll 0 - 1) \kket{\Psi} = (\tilde{L}\ll 0 - 1) \kket{\Psi} &=& L\ll n
\kket{\Psi} = \tilde{L}\ll n\kket{\Psi} = 0,  
\nn
%\nn \\
% n &\geq& 1\ .
\nn
\eee
with $n \geq 1$. The resulting theory of the 
$X\uu\m $ coordinates has the status of an effective theory only, in that it should be
thought of as an expansion that is valid in the limit where the physical length
of the string is much larger than $\sqrt{\apr}$.  
%This does not stand in contradiction with the property that 
%the dynamics of the $X$ fields should be exactly conformal in the two-dimensional sense; we apply the
%effective string theory only in the limit where the conformal
%invariance is spontaneously broken by a classical solution describing a long string.

The introduction of the PS term may appear somewhat ad hoc, but
it has been shown \cite{Polchinski:1991ax, natsuume, Aharony:2010db} that the structure of this theory emerges from
critical string theory when the latter assumes a geometry that is holographically dual to a
gauge theory that confines in the infrared.   Such a geometry is generically a warped product of
Minkowski space with a fifth holographic direction $\phi$.  This framework exhibits a
negative potential energy in five dimensions, and a warp factor for Minkowski space that assumes a
finite global minimum at some $\phi = \phi_0$.  When one considers a string with a large
physical length in the four Minkowski directions $X\uu\m$, localized at $\phi = \phi\ll 0$
in the holographic direction, the fluctuations $\hat{\phi} \equiv \phi - \phi\ll 0$ receive large
masses in the two-dimensional sense, and can thus be integrated out.
From this starting point, the PS term is an effective interaction generated by integrating out 
$\phi$ and the other degrees of freedom.  It is
in this way that effective string theory is related to critical string theory:  The former emerges from the latter
as a Wilsonian low-energy effective theory, under certain well-defined 
conditions.

\section{Structure of the calculation}
%The essential building blocks are correlators computed in the effective CFT of the $X$ fields.  
The underlying treatment is similar in structure to the
theory of a straight, static string, as in \cite{Polchinski:1991ax}, 
but the rotating state in which we evaluate correlation functions is somewhat
more complicated.  
To simplify the calculation, we consider the special case of the
leading Regge trajectory, composed of 
lowest-mass states for given values of the angular momentum.
This simplifies the calculation in three ways.

First, the physical state conditions with $n\geq 1$ are all automatically satisfied for the lowest-energy state
carrying a fixed set of Noether charges (here, Poincar\'e generators $P\uu\m$ and $J\uu{\m\n}$, and
$L\ll 0 - \tilde{L}\ll 0$), so long as the Noether generators are
exactly conserved and the lowest state with those charges is unique.  
These conditions are satisfied in the effective string state 
with the kinematics we consider.  Then, the remaining
physical state condition, specified by $L\ll 0$, dictates 
the first-order shift in the mass-squared of the string state.  The
calculation of $\Delta M\sqd$ at order $J\uu 0$ then reduces 
to a first-order shift in the eigenvalue of the 
worldsheet Hamiltonian.  By the usual methods of first-order 
perturbation theory, this shift in the worldsheet energy $E_{\rm ws}$ reduces 
to an expectation value of the interaction Hamiltonian $\hat{H}\ll{ \rm first-order}$ in the 
free-field state with Noether charges $P\uu\m$ and $J\uu{\m\n}$.  
Schematically, we have
\bbb
\Delta M\sqd \big |\ll{ { { \rm closed} \atop { \rm first-order } } } &=&  { 2\over{\apr}} \cc \Delta E\ll{\rm ws} \big |\ll{ \rm first-order} \ ,
\nn \\
\Delta M\sqd \big |\ll{ { { \rm open} \atop { \rm first-order } } } &=&  {1\over{\apr}} \cc \Delta E\ll{\rm ws} \big |\ll{ \rm first-order} \ ,
\nn\\
\Delta E\ll{\rm ws} \big |\ll{ \rm first-order} &=& \bbra {(P,J)}\ll{\rm free} \hat{H}\ll{ \rm first-order} \kket{(P,J)}\ll{\rm free}\nn \ ,
\eee
where we have assumed the standard coordinate periodicity $\s\uu 1 \simeq \s\uu 1 + 2\pi$.

Second, the expectation value of an operator in an eigenstate of large
angular momentum $J$ is approximated at leading order by the classical
value of that operator in a rotating solution with a helical
symmetry (i.e.,~a symmetry under a combination
of shifting $\s\uu 0$ and $X\uu 0$, and rotating the spacelike coordinates under $SO(D-1)$).  The corrections
to the leading-order value are calculable and of order ${1/ J}$ or smaller, relative to the classical value.  These
corrections can be computed in a straightforward manner (for instance, by representing definite-$J$ states as contour intervals
of coherent states, and evaluating expectation values in a saddle point expansion at large $J$).  The interaction
Hamiltonian is of order $\b |X|\uu 0$, and therefore exhibits a classical value of order $\b |J|\uu 0$ in eigenstates of large $J$.
The corrections to the classical value are of order $|J|\uu{-1}$ at most, and therefore do not contribute to $\Delta M\sqd$ at
order $J\uu 0$.  Thus, the order-$J\uu 0$ shift of the mass-squared is proportional to the order-$J\uu 0$ shift
in the eigenvalue of the worldsheet Hamiltonian, which is simply the
classical value of the interaction Hamiltonian at
that order.  In other words, $\bbra {(P,J)} \ll{\rm free}
\hat{H}\ll{ O(\b\uu 1)} \kket{(P,J)}\ll{\rm free}$ is the classical
value of $H\ll{ O(\b\uu 1)}$ on the unperturbed rotating solution,
plus corrections of $O(|J|\uu{-1})$. 
%\bbb
%  \bbra {(P,J)} \ll{\rm free} \hat{H}\ll{ O(\b\uu 1)} \kket{(P,J)}\ll{\rm free} ~~~~~~~~~~~~~~~~~~  & ~ &
%\\ \nn
% =  ({\rm classical~value~of~}H\ll{ O(\b\uu 1)}~{\rm in~unperturbed~rotating~solution}) & &
% \\ \nn
% + O(|J|\uu{-1})\ .  ~~~~~~~~~~~~~~~~~~~~~~~~~~~~~~  & ~ &
%\eee
We will describe the free Fock states $\kket {(P,J)}\ll{\rm free}$ and the corresponding helical solutions explicitly below.

The third simplification is that we need not 
use the explicit form of the interaction Hamiltonian, even
for purposes of classical evaluation.  By
straightforward manipulations in classical mechanics, it is possible to show that the first-order shift in the energy of
the lowest classical solution with fixed Noether charges 
is equal both to the value of the interaction Hamiltonian, 
and equivalently to the negative of the value of the interaction 
Lagrangian, evaluated in the unperturbed (zeroth-order) helically
symmetric solution with the appropriate Noether charges:
\bbb
\Delta E\ll{{\rm ws}} \big |\ll{ O(\b\uu 1)} = - \int \cc d\s\uu 1 \cc
       {\cal L}\ll{\rm PS} \big |\ll{\rm rotating~solution}\ .
\eee

By restricting our attention to 
lowest-mass string states of fixed $J$, and using these three simplifications, we reduce
the calculation of the order-$|J|\uu 0$ term in the mass-squared of
the string state to computing the free Casimir energy, minus the 
classical value of the interaction Lagrangian in the helically
symmetric solution.
Quantum contributions involving the interaction Lagrangian are at most of order $|J|\uu{-1}$.
The same applies to any effect of non-universal terms in the worldsheet action, which (with the exception of a
single boundary operator, which we shall discuss below) scale with
negative powers of $X$ (and therefore of $|J|$) 
when contributing to worldsheet energies, even at the classical level.  
The result is that we need only take into account the one-loop free-field Casimir energy and the
classical on-trajectory PS term.  These contributions simply add.  
\section{Closed Strings}
For closed string configurations in $D\geq 5$, 
with rotation in two planes, the parametric scaling of corrections in our analysis mostly
corresponds to that of the large-$R$ expansion
in the case of a straight, stretched string 
\cite{Polchinski:1991ax,Drummond:2004yp,Drummond:2006su,Dubovsky:2012sh,Dubovsky:2012wk,Dubovsky:2013gi,Aharony:2013ipa,Aharony:2009gg,Aharony:2010cx,Aharony:2010db,Aharony:2011ga,Aharony:2011gb,AharonyUnPub}, under the replacement $ R \to \sqrt{J\apr}$.  
This is expected, because the physical length of the rotating string is proportional to $\sqrt{J\apr}$.  The main difference
between the rotating case and the static case is that the PS term \rr{psterm} makes a contribution to the energies
of large strings at first subleading order in the inverse size of the string, which does not happen in the static case.  This is not mysterious:
It is attributable to the simple fact that the classical value of the PS Lagrangian is zero for the straight static string and
nonzero for the rotating string.

We start by describing the lowest-energy state with the quantum numbers of interest in the free theory.
Generally, the lowest-energy
eigenstates of angular momentum in the $Z,\ \Zb$ plane are those generated by acting $J$ times with
left- or right-moving creation operators, with one unit of $L\ll 0$ or $\tilde{L}\ll 0$ each.  
For the closed string, there is also the additional restriction of level-matching, meaning that the number
of left- and right-moving creation operators must be the same.  

The states of interest are such that, in
an appropriately chosen Cartan decomposition, the angular momenta are aligned
with the ``3'' direction of the self-dual and anti-self-dual
$SU(2)_\pm$ subgroups of the $SO(4) \subset SO(D-1)$ little group.
For $D\geq 5$, we can consider general angular momenta in both planes, 
where the string states are chosen to be primaries of both
$SU(2)_\pm$ subgroups, and the total angular momentum quantum numbers 
are $J\ll\pm \equiv \hh (J\ll 1 \pm J\ll 2)$ in $SU(2)\ll\pm$.
In other words, minimizing the energy over highest-weight vectors of $SU(2)\ll + \times SU(2)\ll -$, with 
total angular momenta $J\ll \pm$ and zero momentum in the $\s\uu 1$ direction,
the unique lowest-energy state in the free theory can be expressed as
%\bbb
%\kket{J\ll + , J\ll - ; ~P} &=& {1\over\sqrt{{\cal N}\upp{{\bf I}}\ll{J\ll +, J\ll -}}}
%\bigg (\a\uu{Z\ll 1}\ll{-1} \a\uu{Z\ll 2}\ll{-2} - \a\uu{Z\ll 1}\ll{-2} \a\uu{Z\ll 2}\ll{-1} 
%\bigg)\uu{J\ll + - J\ll -} 
%\nn\\
%&& \times \bigg (   \a\uu{Z\ll 1}\ll{-1} \cc   \bigg )\uu{2J\ll -} \cc \kket{0\ ; ~P} \ ,
%\eee
%for the CFT on the interval, and 
\bbb
\kket{J\ll + , J\ll - ; P}\ll{\rm free} &=& {1\over\sqrt{{\cal N}\upp{{\bf S\uu 1}}\ll{J\ll +, J\ll -}}} \left( 
 \a\uu{Z\ll 1}\ll{-1} \cc \tilde{\a}\uu{Z\ll 2}\ll{-1}
 -
  \a\uu{Z\ll 2}\ll{-1} \cc \tilde{\a}\uu{Z\ll 1}\ll{-1}
  \right )\uu{J\ll + - J\ll -} 
  \nn\\         % comment these two lines in for 2-column version
&&  \times
  \left (   \a\uu{Z\ll 1}\ll{-1} \cc \tilde{\a}\uu{Z\ll 1}\ll{-1}  \right )\uu{J\ll -} \cc \kket{0\ ; P}\ll{\rm free}\ .
\eee 
%for the CFT on the circle.  
The energy under the free-field Hamiltonian is
\bbb
&&E\ll{\rm ws}\upp{{\rm free}} =  \left({1\over 2} \cc \apr P\sqd + 2\cc J\ll + \right)  - {D\over{12}}\ ,
%\tilde{L}\ll 0\upp{\rm free}= L\ll 0\upp{\rm free} &=& \apr P\sqd + 3J\ll + - J\ll - \ .
%\nn\\
%&=& \apr P\sqd + {\max}(|J\ll{1}| , |J\ll{2}|) + 2 \cc {\min}(|J\ll{1}| , |J\ll{2}|)\ .\quad
\eee
where the last term is the usual free-field Casimir energy. %\footnote{Note that for generic 
%angular momenta in two planes, the oscillator energies of the lowest states
%of the CFT with and without boundary are not proportional, and are not related to one another by the usual doubling
%trick; that relation is special to solutions with angular momentum in a single plane.}
%\be
%L\ll 0 &=& \apr P\sqd + 3J\ll + - J\ll - 
%\nn\\
%&=& \apr P\sqd + {\max}(|J\ll{1}| , |J\ll{2}|) + 2 \cc {\min}(|J\ll{1}| , |J\ll{2}|)\quad
%\eee
%on the interval and
%\bbb
%L\ll 0 \upp{\rm free}= \tilde{L}\ll 0\upp{\rm free} = {1\over 4} \cc \apr P\sqd + J\ll + \ .
%\nn\\
%  =  \tilde{L}\ll 0 = {1\over 4} \cc \apr P\sqd + \hh (J\ll{1} +  J\ll{2})\ .
%\eee
The values of the normalization constants of the Fock states (${\cal N}\upp{{\bf S\uu 1}}\ll{J\ll +, J\ll -}$) 
will of course drop out of all observables, but we can choose them so
that the Fock states $\kket{J\ll + , J\ll -;P}_{\rm free}$ of the
nonzero modes are unit-normalized.  Furthermore, we can choose that
$J_+ > J_-$, for instance, without loss of generality.

Now we wish to evaluate expectation values of operators (for instance, the interaction Hamiltonian) in
this state.  At large $J$, we expect that the quantum state is approximated
by a classical solution, and indeed this is the case.
Expectation values of operators in $\kket{(J,P)}$ are given to leading
order in $J$ by the classical values of those operators in a particular
helically symmetric solution, suitably averaged over rotations in the $Z\ll{1,2}$ planes
if the operator in question is not already symmetric.  

A general helically symmetric classical solution is given by
\bbb
X\uu 0 &=& \apr\cc P\uu 0 \cc \s\uu 0\ ,
\xxn
Z\ll i &=& - i \cc \sqrt{{{\apr}\over 2}} \cc \left(\a\uu{Z\ll i}\ll{-1} e^{i \s\uu + } 
+ \tilde{\a}\uu{Z\ll i}\ll{-1} e^{i \s\uu - } \right)\ ,
\xxn
\Zb\ll i &=&  i \cc \sqrt{{{\apr}\over 2}} \cc \left( \a\uu{\Zb\ll i}\ll 1 e^{- i \s\uu + } 
+ \tilde{\a}\uu{\Zb\ll i}\ll 1  e^{- i \s\uu - } \right)\ ,
\label{closedStringClassicalSolutions}
\eee
with $i \in \{1,2\}$, and with the mode amplitudes obeying $(\a\uu {Z\ll i}\ll {-n})^* = \a\uu{\bar{Z}\ll i}\ll n$.
%The spacetime energy $P\uu 0$ is
%\be
%P^0 =  \frac{1}{2\pi\apr}\int d\s^1 \pp_0 X^0 \ .
%\ee
%It is convenient to define
%\bbb
%\a\uu{\Zb\ll i}\ll 1 = \sqrt{2} \cc \ell\ll i \ , &\quad& \tilde{\a}\uu{\Zb\ll i}\ll 1 = \sqrt{2} \cc \tell\ll i\ ,
%\xxn
%\a\uu{Z\ll i}\ll{-1} = \sqrt{2} \cc \ellba\ll i\ , &\quad& \tilde{\a}\uu{Z\ll i}\ll {-1} = \sqrt{2} \cc \tellba\ll i\ ,
%\een{littleLDefs}
In terms of the modes, the free angular momentum generators are:
\be
J\ll{i} &=& \frac{i}{4\pi\apr}\int d\s\uu 1 \left( Z\ll i \dot{\bar
Z}\ll i - \bar Z\ll i \dot Z\ll i \right)  
\nn\\
&=&  \hh |\a\uu{Z\ll i}\ll{-1}|^2 +\hh |\tilde{\a}\uu{Z\ll i}\ll{-1}|^2\ .\nn
\ee
Fixing $P\uu\m$, $L\ll 0 - \tilde{L}\ll 0 = 0$, and the values of the angular momenta $J\uu{\m\n}$, and then
choosing values of the Fourier coefficients
that minimize $L\ll 0 + \tilde{L}\ll 0$ , we find 
\bbb
&& \a\uu{Z\ll 1}\ll{-1} =  \a\uu{\Zb\ll 1}\ll{1} =  \tilde{\a}\uu{Z\ll 1}\ll{-1} =  \tilde{\a}\uu{\Zb\ll 1}\ll{1} =   \sqrt{J\ll 1}\ ,
\xxn
&&\a\uu{Z\ll 2}\ll{-1} =  \a\uu{\Zb\ll 2}\ll{1} =  -\tilde{\a}\uu{Z\ll 2}\ll{-1} =  -\tilde{\a}\uu{\Zb\ll 2}\ll{1} =   \sqrt{J\ll 2} \ .
\een{RepresentativeClosedSolutionApprox}
%In the free theory of the massless $X\uu\m$ degrees of freedom, we construct eigenstates of
%$P\ll\m$ in the standard way, as eigenstates of the conjugate momentum to the zero mode of $X\uu\m$.  
%We are considering massive string states, so we can immediately move to the center of
%mass frame, in which $P\uu 0$ is nonzero and the other $P\ll i$ are vanishing, including $P\ll z$ and $P\ll\zb$.  
Evaluated in this rotating solution, the contribution of the PS anomaly term, evaluated
in the rotating ground state, takes the form 
\be
{\cal L}_{{\rm PS}\atop{\rm rotating~solution}} = - {{\b J\ll -\sqd}\over{2\pi\sqd}} \cc 
{{{\rm sin}\sqd( 2  \s\ll 1 ) }\over{(J\ll + - J\ll -
    \cc {\rm cos}( 2  \s\ll 1   ))\sqd}}\ .
\ee
This Lagrangian density
becomes singular at the endpoints $\s\ll 1 = 0$ and $ \pi $, in the limit $J\ll + = J\ll -$.  
This limit is imposed automatically in $D=4$, as the little group $SO(D-1)$ has rank one, and $J_2$ must vanish.
The singularity corresponds to the development of a fold in the
string, and we defer a careful treatment of these cases for future work. 

However, the integral is finite for generic angular momenta in $D\geq 5$, 
and we do not need to regulate or renormalize.  
The resulting value of the mass-squared, to order $J\uu 0$, is 
\bbb
&&\kern-30pt  M\sqd\ll{{\rm closed}} 
 =  {1\over\apr} \cc \Biggl[
 2 (J\ll 1 +J\ll 2) - {{D-2}\over 6}
\nn\\          % comment in for 2-col version
&&\kern-10pt + {\frac{26-D}{12}}  \cc \left(  \left(\frac{J_1}{J_2}\right)\uu{{1\over 4}}
- \left(\frac{J_2}{J_1} \right)\uu{{1\over 4}} \right)^2  \Biggr] + O(J^{-1})\ .
\label{finalClosedStringSpectrum}
 \eee
The contribution from the PS term is nonzero unless $J\ll 1 = J\ll 2$, or $D=26$.

%------------------------------------------------------------------------------
\section{Open Strings}
%------------------------------------------------------------------------------
For open rotating strings with Neumann boundary conditions, a qualitatively new effect will appear:  
The anomaly term \rr{psterm} in the effective action becomes singular
at the boundary, and the singularity is non-integrable. 
%the integrated value of the classical Lagrangian in the rotating
%solution diverges. % repetitive
This divergence is a short-distance singularity, which can be removed by regularization and
renormalization.  In particular, we can remove the divergence by
adjusting the coefficient of a boundary 
operator -- the unique marginal boundary operator with the correct $X$-scaling to cancel the divergence.

In terms of $J_\pm$, the lowest-energy, highest-weight states in the free theory are specified by
\bbb
\kket{J\ll + , J\ll - ; P}\ll{\rm free} &=& {1\over\sqrt{{\cal N}\upp{{\rm open}}\ll{J\ll +, J\ll -}}}
\left (\a\uu{Z\ll 1}\ll{-1} \a\uu{Z\ll 2}\ll{-2} - \a\uu{Z\ll 1}\ll{-2} \a\uu{Z\ll 2}\ll{-1} 
\right)\uu{J\ll + - J\ll -} 
\nn\\          % comment in for 2-col version
&& \times 
\left (   \a\uu{Z\ll 1}\ll{-1} \cc   \right )\uu{2J\ll -} \cc \kket{0\ ; P}\ll{\rm free} \ ,
\eee
where ${\cal N}\upp{{\rm open}}\ll{J\ll +, J\ll -}$ is again a
normalization constant.  
The energy under the free-field Hamiltonian now appears as
\bbb
&&E\ll{\rm ws}\upp{{\rm free}} =  \apr P\sqd + 3J\ll + - J\ll -  - { D\over{24}}\ .
\eee

Analogous to the closed string, 
expectation values of rotationally symmetric operators in this state are given to leading
order in $J$ by the classical values of those operators in a particular
helically symmetric classical solution, which minimizes the energy for its Noether charges, and
takes the form 
\bbb
X\uu 0  & = &  2 \apr P^0 \s^0 
\xxn
\Zb\ll 1 &=&   i \sqrt{\frac{\apr}{2}}  \a_1^{\bar Z_1}  \biggl ( \cc
 e^{-i  \s\uu + }  + 
 e^{-i  \s\uu - }
    \biggl )\ 
    \xxn
    \Zb\ll 2 &=&   i \sqrt{\apr\over 2} \frac{\a_2^{\bar Z_2}}{2}  \biggl ( \cc
 e^{-2 i  \s\uu + }  + 
 e^{-2 i  \s\uu - }
    \biggl )\ 
\xxn
Z\ll 1 &=&   -i \sqrt{\apr\over 2}  \a_{-1}^{Z_1}  \biggl ( \cc
 e^{i  \s\uu + }  + 
 e^{i  \s\uu - }
    \biggl )\ 
    \xxn
    Z\ll 2 &=&   -i \sqrt{{{\apr}\over 2}} \frac{\a_{-2}^{Z_2}}{2}  \biggl ( \cc
 e^{2 i  \s\uu + }  + 
 e^{2 i  \s\uu - }
    \biggl )\ ,
\eee
%If we define the variables
%\be
%\a_1^{\bar Z_1} = \sqrt{2} \ell\ll 1  &\quad & \a_{-1}^{Z_1} = \sqrt{2} \bar\ell\ll 1 
%\nn\\
%\a_2^{\bar Z_2} = 2 L\ll 2  &\quad & \a_{-2}^{Z_2} = 2 \bar L\ll 2 \ ,
%\ee
%then 
%the classical zeroth-order Virasoro constraints give
%\bbb
%\apr (P\uu 0)\sqd = |\ell\ll 1|\sqd +2\cc |\BIGell\ll 2|\sqd \ .
%\een{DispRelnOpenAmps}
%and the
%the lowest-energy solution is 
with
\bbb
\a_1^{\bar Z_1} = \sqrt{2J\ll 1}   &\quad & \a_{-1}^{Z_1} = \sqrt{2 J\ll 1} 
\nn\\
\a_2^{\bar Z_2} = 2 \sqrt{J\ll 2}  &\quad & \a_{-2}^{Z_2} = 2 \sqrt{J\ll 2} \ .
\eee
%\bbb
%|\ell\ll 1| = \sqrt{ J\ll 1}\ , \llsk |\BIGell\ll 2| = \sqrt{J\ll 2}
%\nn \\  \\
%\ell\ll 2  =  \BIGell\ll 1 = 0\ . \nn
%\eee
%and the leading-order dispersion relation in terms of the angular momenta $J\ll\pm$ 
%then appears as
%$(P\uu 0)\sqd \simeq {{(3 J\ll + - J\ll - )}/{\apr}}$.
%
As above, evaluation of expectation values of operators in this quantum state of the free theory
gives answers to leading order in $J$, equal to the classical values
of those operators evaluated on the solution above (and, if necessary, averaged over rotations
of $Z\ll{1,2}$, when the operator is not rotationally symmetric from
the outset).

As noted, the PS term for open strings exhibits a short-distance 
divergence near the boundaries, which can be canceled with an
appropriate boundary counterterm, and
an analysis of scale-invariant boundary operators consistent with Lorentz 
symmetry reveals that only one such independent operator is available to regulate
the divergence.  
%\redd{In particular, all Lorentz-invariant operators of
%conformal dimension one and non-negative $X$-scaling, on the
%constraint space of a vanishing stress tensor, the Virasoro conditions,
%equations of motion, Neumann boundary conditions, and vanishing total
%derivatives in the $\s_0$ direction, reduce either to zero, or to a single
%independent operator, which corresponds to the effect of a quark
%mass.  DEPT OF REDUNDANCY DEPT. (WE SAY THIS BELOW, WITH BETTER LANGUAGE)}  
% We have verified this, and there is a description in the longer
% draft.  
In particular, the short-distance divergence can be shown to take
the form of a quark-mass boundary operator, with a coefficient that
diverges as some short-distance regulator scale $\epsilon$ is taken to
zero.  The divergence can thus be cancelled with a corresponding 
counterterm. 
%\redd{in the form of a quark-mass boundary operator.}   
%This agrees with the
%quasi-static-gauge analyses performed, for example, in \cite{Baker:2000ci,Baker:2002km} 
%in a $4D$ effective string theory, and in \cite{Kruczenski:2004me} in a holographic model.  

We now compute the value of the regulated classical action, and renormalize it with a boundary
counterterm to extract the finite piece that 
contributes to the Regge intercept.  
To regulate the divergence, we modify the form of the PS Lagrangian to cut off the
singular behavior of the integrand:
\be
&&{\cal L}_{\rm PS} = {\frac{\beta}{2\pi}} {\frac{ ( \pp\ll + \sqd X \cdot \pp\ll - X ) ( \pp\ll +  X \cdot \pp\ll -\sqd  X )  }
			{(\pp\ll + X \cdot \pp\ll - X)\sqd  }}
			\nn \\ & &  \nn \\ 
&&\to {\cal L}_{\rm PS,~reg} \equiv {\frac{\beta}{2\pi}} {\frac{ ( \pp\ll + \sqd X \cdot \pp\ll - X ) ( \pp\ll +  X \cdot \pp\ll -\sqd  X )  }
			{(\pp\ll + X \cdot \pp\ll - X)\sqd  + \apr\cc \e\uu 4 \cc (\pp\ll +\sqd X \cdot \pp\ll - \sqd X)  }}\ , \nn
			\ee
where the coefficient $\apr\e\uu 4$ of the regulating term is chosen for later convenience.  This form
of the modification preserves $D$-dimensional Poincar\'e invariance and
all other symmetries of the theory, including two-dimensional scale invariance.  (These symmetries will drastically
constrain the form of the available counterterms.)

Next, we must calculate the $\s_1$ integral of the classical value of
${\cal L}_{\rm PS}$, from $0$ to $\pi $, up to and including order $\epsilon\uu 0$.  Before extracting
the finite term, we first consider the form of the divergence as we send $\e\to 0$.  
To this end, we introduce a new integration variable $u$ by
  rescaling $\s_1$:
\bbb
\s\uu 1 = \e\cc \langle{\cal O}\lp{{\rm quark}}\rangle\cc u\ ,
\eee
where ${\cal O}\lp{{\rm quark}}$ is the boundary operator
\bbb
{\cal O}\lp{{\rm quark}} \equiv (\pp\ll {\s\uu 1}\sqd X\cdot \pp\ll {\s\uu 1}\sqd X)\uu{{1/4}}\ ,
\eee
and $\langle{\cal O}\lp{{\rm quark}}\rangle$ is its value in the classical rotating solution, proportional to $(J_ 1 + 8 J_2)^{1/4}$.
Expanding the integrand $d\s\uu 1 \cc{\cal L}_{\rm PS,~reg}$ in terms of $u$,
we see that the $\epsilon^{-1}$ divergence is proportional, with a universal
coefficient, to $\langle{\cal O}\lp{{\rm quark}}\rangle$.  We conclude that the divergence of the PS action near a Neumann boundary
can be renormalized with a boundary counterterm proportional to $\e\uu{-1} {\cal O}\lp{{\rm quark}}$.

The boundary operator ${\cal O}\lp{{\rm quark}}$ is quite interesting
in its own right.  It is the only Lorentz-invariant, independent
  boundary operator in the theory with
marginal scaling dimension and non-negative $X$-scaling.  
That is, all other linearly independent 
boundary operators of dimension one respecting $D$-dimensional Poincar\'e symmetry
can be eliminated by field redefinitions, Virasoro constraints, and by
discarding 1) total derivatives tangent to the boundary, and 2) operators with negative
$X$-scaling (and therefore negative $J$-scaling).   
In fact, this operator can be added to the boundary action with
a finite coefficient, which adjusts the mass-squared for the open string by $\Delta M\sqd \propto
J\uu{{1/4}}$ at large $J$.  This $J\uu{{1/4}}$ term has been studied in noncovariant gauges
(see, e.g.,~\cite{Baker:2002km, Wilczek:2004im, Kruczenski:2004me}), and corresponds to the effect
of a finite mass for a quark at the string endpoint.

We now demonstrate the renormalizability of the divergence
directly, and extract the finite term by performing
the integral.  One way to carry this through is to invoke 
a change of variables $w \equiv \expr{2 i \s\ll 1}$, 
taking the contour on the unit circle $|w|=1$.  The integrand is a rational function of $w$, and we can
evaluate the integral by taking residues.  The divergent terms come from four poles which
approach $w=1$, two from each side of the unit circle, which give rise to the $\e\uu{-1}$ behavior of the integral
in the $\epsilon\to 0$ limit.  
%\footnote{Note that the required contour is precisely the
%unit circle;  there are singularities clustering near 
% $w = 1$ from both sides of the unit circle, which give rise to the UV divergence in the $\eha\to 0$ limit.}
In terms of $\s\uu 1$, these are boundary contributions, corresponding to the endpoints
of the string.  Together, they contribute 
\be
\Delta M^2_{\rm open} = \frac{1}{\epsilon}\, \frac{26-D}{24 \apr}
\left(J_1 + 8 J_2\right)^{1/4}  + ({\rm finite})\ .
\ee
%where $\e^4 \equiv \csh^2 / \apr$.
In addition, there are contributions from poles in the interior of the unit circle in the $w$-plane.  These are finite in the
$\e\to 0$ limit.
After renormalizing the $\e\uu{-1}$ divergence by adding our 
%$J\eha\uu{-1} J\ll 1\uu{{1\over 4}}(1 + 8a)\uu{{1\over 4}}$ term with a 
counterterm to the boundary Lagrangian proportional 
to $(\pp^2_+ X \cdot \pp^2_- X)^{1/4}$, we obtain
\bbb
&&\kern-00pt M\sqd\ll{{\rm open}}  
= \frac{1}{\apr}\Biggl[ {J\ll 1 + 2 J\ll 2  - {{D-2}\over{24}}}
\nn\\          % comment in for 2-col version (and fix \left[ and \right]) 
&&\kern+10pt + \frac{26-D}{24} \left( 
    -4 + {{3\cc J\ll 1 + 4  \cc J\ll 2 }\over{J\ll 1\uu{\hh}\cc \sqrt{J\ll 1 + 8 J\ll 2}}}  \cc \right) \Biggr]  
    + O(J\uu{-1})\ .
\nn\\
&&
%   \xxn
%    = \frac{(1 + 2 a) J\ll 1}{\apr}
%    + {1\over {24\cc\apr}}\left[ \left( 3 - {{3 + 4a}\over{\sqrt{1 + 8a}}} \right) \cc D 
%    + \left( -102 + {{26(3 + 4a)}\over{\sqrt{1 + 8a}}}
%    \right) \right]\ .
%    \xxn
%&&
\label{finalOpenStringSpectrum}
\eee
For angular momenta lying in a single plane (i.e., when $J_2 = 0$), the mass-squared equals 
$M\sqd\ll{{\rm open}} = (J\ll 1 - 1)/\apr$, independent
of $D$.  
%As noted above, $J\ll 2$ must vanish when $D=4$, so here
%the asymptotic Regge intercept is exactly the same as that exhibited by 
%microscopic bosonic string theory in the critical dimension $D=26$.  
Of course, when $D=26$, we obtain $M\sqd\ll{{\rm open}} = (J\ll 1 + 2 J\ll 2 - 1)/\apr$.  
This is the case in which the bosonic 
string theory is well-defined microscopically, and the singular PS
anomaly term is absent.  

It is worth emphasizing that
we have fine-tuned the coefficient of the quark mass operator ${\cal O}\ll{{\rm (quark)}}$ so that there is no
term of order $J\uu{{1/ 4}}$ in the mass-squared formula
\rr{finalOpenStringSpectrum}.  In other words,
we should generally expect a $J\uu{{1/ 4}}$ term in the open-string mass-squared, unless the mass of the
quark at the endpoint is light compared to the scale of the string tension.
\nopagebreak
\section{Relation to Other Work}
Our results should be compared to calculations of the subleading large-$J$ corrections to the
effective string spectrum in the existing literature.  
The earliest work known to us on the subject appears in 
references \cite{Baker:2000ci,Baker:2002km}, which report subleading
corrections to the mass-squared of the open string that differ from
our results at order $J\uu 0$, implying a different value for
the asymptotic Regge intercept.  In particular,
these papers analyze the special case $J\ll 2 = 0$, and quantize the Nambu-Goto
action directly in a version of static gauge,  
using only the induced metric $\bar{g}_{ab} \equiv \partial_a X\uu\m \cc 
\partial_b X\ll\m$ to define the theory classically, as well as to regulate and renormalize it at the quantum level.  They
report a value for the mass-squared of the open string equal to eq.~\rr{finalOpenStringSpectrum},
except with the term proportional to $D-26$ (corresponding to the contribution of the PS term in 
the conformal-gauge calculation) absent.  The same
value of the asymptotic intercept for the rotating Nambu-Goto string quantized in static gauge
has appeared elsewhere in
the literature (e.g.,~in \cite{Kruczenski:2004me}).
%\nopagebreak

While a thorough review of the results in the existing literature is beyond the scope of
this letter, we comment on the origin of one potential source of
disagreement between the static-gauge 
calculations of the mass-squared and our own covariant-gauge calculations.
In the Nambu-Goto string,
Weyl symmetry is not a
gauge symmetry, and there is no need for an anomaly-cancelling term 
in the action, as the theory is defined using
only the induced metric $\bar{g}\ll{ab}$.  If the theory is quantized 
consistently in this framework, gauge-invariant 
observables must match those computed in any other gauge.  
We note, however, that certain quantities involving the induced metric 
$\bar{g}_{ab}$, such as the determinant of the scalar Laplacian 
$\bar{\nabla}\sqd \equiv \nabla_a \bar{g}^{ab} \nabla_b$,  are vulnerable to subtle errors, due to 
the dependence of the the quantum effective action on the background 
classical solution, through the role of the induced metric
$\bar{g}_{ab}$ in the renormalization of the theory.
We obtain a result for the renormalized determinant of the Laplacian of the induced metric, for instance, that
%\nopagebreak
differs from the value stated in equation $(9.8)$ of \cite{Baker:2002km}.\footnote{In computing the
determinant, we use a standard formula
for the renormalized determinant of the Laplacian of a conformally
flat metric, e.g.,~formula (3.4.18) of \cite{Polchinski:1998rq}, with
the value $a\ll 1 = - {1/ {12}}$ 
for a single scalar field, and taking the conformal factor $\expr{2\o}$
to be that of the induced metric $\bar{g}_{ab}$, so 
that $\o = \hh \cc{\rm ln}(\pp\ll + X \cdot \pp\ll - X)$. This formula 
for the determinant gives an additional contribution to the quantum
effective action proportional to the integral of $(\nabla \o)\sqd$.  Summing
over the $D-2$ transverse coordinates, we find a value for the
quantum effective action differing from the results of \cite{Baker:2002km}
by an amount $\Delta {\rm ln}(Z) = i\int \cc d\s\uu 0 \cc d\s\uu 1 \cc \Delta {\cal L}\ll{\rm qu.}$, where
\bbb
\Delta {\cal L}\ll{\rm qu.} \equiv -{{D-2}\over{24\pi}}\cc {\frac{ ( \pp\ll + \sqd X \cdot \pp\ll - X ) ( \pp\ll +  X \cdot \pp\ll -\sqd  X )  }
			{(\pp\ll + X \cdot \pp\ll - X)\sqd  }}\ .\nn
\eee
Note that the $D$-dependence exactly reproduces that of eqns.~\rr{psterm}
and \rr{betaval} above.
(We thank Shunsuke Maeda 
and Jonathan Maltz for a detailed discussion of this determinant.)
We expect that the measure terms from the gauge-fixing of the
diffeomorphism symmetry of the path integral in \cite{Baker:2000ci,Baker:2002km} 
may produce contributions of the same form, with $D$-independent coefficients.
}
We anticipate that a careful recalculation the one-loop contributions to the intercept
in the Nambu-Goto string in static gauge would make up the difference
between the existing values in the literature and the covariant-gauge result presented in this letter. 
\vskip-.3in
%%-----------------------------------------------------------------------------------------------------------------
%%-----------------------------------------------------------------------------------------------------------------
\section{Discussion}
%%-----------------------------------------------------------------------------------------------------------------
%%-----------------------------------------------------------------------------------------------------------------
%We have developed the tools necessary to
%compute subleading contributions to the mass spectra of confining strings
%in the Polchinski-Strominger covariant effective string theory, in the expansion in inverse angular momentum at first subleading order, applying
%these tools to the simple case of the leading Regge trajectory in bosonic planar QCD.  We have
%used the effective string theory approach of
%Polchinski and Strominger \cite{Polchinski:1991ax}, which preserves conformal invariance
%order-by-order in our semiclassical expansion around the rotating string
%solutions corresponding to different ratios of angular momenta in two planes.  We calculated the
%order $J\uu 0$ correction to the mass-squared, for closed
%strings in $D\geq 5$ with generic angular momenta in two planes, and for open strings with Neumann boundary
%conditions in $D \geq 4$. 
The results above (in particular, eqn.~\rr{finalOpenStringSpectrum})
constitute the first step toward using the covariant formalism to connect the
higher-resolution predictions of effective string
theory with experiments.   
However, the value of
the Regge intercept computed here for open strings with bosonic endpoints is notably different from the intercept for
the best-fit trajectory to data in the actual hadron spectrum
(see, e.g., \cite{someExperimentalResults}, or \cite{PDG} for the latest underlying data).  
Where we find an
asymptotic Regge intercept for the open string of $\a\ll 0 = -M_0^2 \apr= 1$ 
on the leading trajectory (for $J\ll 2 = 0$ and $D=4$), the data indicate an asymptotic intercept in
real QCD that
is roughly half of this value for the trajectory of the $\rho$ meson.
It is an important problem to understand what effects might be
needed to eliminate this difference.  Estimates of corrections from $O(|X|)^{-2}$
terms in the action, quark masses, and electromagnetic corrections all appear to
be too small.  Regarding the latter, we note that
electromagnetic interactions between quarks must scale as $\Delta P\uu 0 \propto \alpha_{em} / L
= \alpha_{em} / \sqrt{J \apr} $, where $L$ is the physical length of
the string.  Thus, these effects contribute to the intercept, but are suppressed by the
fine-structure constant $\alpha_{em}$.

Nonplanar effects may contribute.  Exchange of singlets between boundaries will be exponentially suppressed
in $\sqrt{J}$ if the lightest singlet is not exactly massless.
However, for approximate chiral symmetry this effect may be large enough to contribute significantly to the
mass squared, even at moderate
$J$.  Furthermore, emission and re-absorption of a singlet by the same boundary may contribute with a power law of $J$.  This
contribution would formally involve the inclusion of degenerate worldsheets into the path integral, so
one would have to understand how to treat the measure near the boundary of the space of smooth
worldsheets in the effective string framework. 

Furthermore, we conjecture that the inclusion of fermonic quarks
carrying quantum numbers of chiral flavor symmetry may 
generate significant contributions at the order of interest, and bring the predictions 
of the effective string calculation closer to the observed value of the intercept on the leading Regge trajectory.
These might be included, for example, by taking a Lagrangian world-line
realization of fermonic quarks (see, e.g.,
\cite{Berezin:1976eg,Frydryszak:1996mu}) and attaching it to the
string endpoints.  We would also expect the incorporation of 
chiral symmetry to set the coefficient of the  $J^{1/4}$ term
naturally to zero, without fine-tuning.  We hope in future work to incorporate these features of planar QCD into the
effective string approach.

The landscape spanned by gauge theory and string
theory is vast, and many of the connections between the two
have yet to be understood in detail in the confining regime, particularly for realistic models such
as nonsupersymmetric planar QCD.
It would represent significant progress toward this goal if
a quantitative theory of the QCD string could indeed make precise contact with the
features of the hadronic spectrum at large $N$ and strong coupling.
%This is rendered particularly true by the relative dearth of reliable
%computational tools in this regime of the dynamics.
%Large-$J$ resonances are states living deeply in the Lorentzian regime
%of nonperturbative QCD.   While the difficulty of
%computing large-$J$ spectra accurately on the lattice, for instance,
%is certain to scale with increasing $J$, the effective string approach becomes simpler and
%more accurate in this limit.

\section{Acknowledgements}
The authors are grateful to many people for valuable discussions, 
including Raphael Flauger, Jacob Sonnenschein, and Taizan Watari.  
We are particularly grateful to Shunsuke Maeda and Jonathan Maltz 
for collaboration on related work, including an analysis of the 
scheme-dependence of the effective string action in various 
noncovariant gauges, and assistance in reconciling the details 
of existing calculations in the literature with the results 
calculated here. The work of S.H. was supported by 
the World Premier International Research Center Initiative, MEXT,
Japan, and also by a Grant-in-Aid for Scientific Research (22740153)
from the Japan Society for Promotion of Science (JSPS).  S.H.~is 
also grateful to the theory group at Caltech for hospitality while this work was in progress.

\end{document}